\documentclass[journal=mamobx,manuscript=article]{achemso}

\usepackage[version=3]{mhchem}
\usepackage[dvipsnames]{xcolor}

\author{Vinay Vaibhav}
\email{vinay.vaibhav@unimi.it}
\affiliation{Department of Physics ``A. Pontremoli", University of Milan, via Celoria 16, 20133 Milan, Italy}

\author{Timothy W. Sirk}
\email{timothy.w.sirk.civ@army.mil}
\affiliation{US DEVCOM Army Research Laboratory, Maryland, United States}

\author{Alessio Zaccone}
\email{alessio.zaccone@unimi.it}
\affiliation{Department of Physics ``A. Pontremoli", University of Milan, via Celoria 16, 20133 Milan, Italy}

\title{Timescale bridging in atomistic simulations of epoxy polymer mechanics using non-affine deformation theory}

\begin{document}

\begin{abstract}
Developing a deep understanding of macroscopic mechanical properties of amorphous systems which lack structural periodicity, has posed a key challenge, not only at the level of theory but also in molecular simulations. Despite significant advancements in computational resources, there is a vast timescale disparity, more than 6 orders of magnitude, between mechanical properties probed in simulations compared to experiments. Using the theoretical framework of non-affine lattice dynamics (NALD), based on the instantaneous normal modes analysis determined through the dynamical matrix of the system, we study the viscoelastic response of a cross-linked epoxy system of diglycidyl ether of bisphenol A (DGEBA) and poly(oxypropylene) diamine, over many orders of magnitude in deformation frequency, below the glass transition temperature. Predictions of the elastic modulus are satisfactorily validated against the non-equilibrium molecular dynamics simulations in the high-frequency regime, and against experimental data from dynamic mechanical analysis at frequencies $ \sim 1 {\rm Hz}$, hence successfully bridging the timescale gap. The comparison shows that non-affine displacements at the atomic level account for nearly two orders of magnitude reduction in the low-frequency elastic modulus of the polymer glass, compared to affine elasticity estimates. The analysis also reveals the role of internal stresses (as reflected in the instantaneous normal modes), which act as to strengthen the mechanical response. 
\end{abstract}

\section{Introduction}

Cross-linked thermosetting polymers and their composites are extensively used in a variety of sectors such as aerospace, construction, and defense equipment, due to their extraordinarily tunable mechanical properties which can be controlled by varying parameters of the network, such as the resin chemistry \cite{dennis2020}, molecular weight \cite{patterson2020, masser2021,dennis2023}, and degree of cross-linking \cite{jang2016}. However, due to complex chemistry and topology, the underlying physics of thermosets is not well described by any single level of theoretical or computational methods \cite{gu2019}. Various atomistically detailed simulation models, properly taking into account the chemical interactions, have been developed in the last few decades and show reasonable rubber and glass properties across the glass transition temperature \cite{sirk2015, elder2015}. Such numerical models have been very successful in explaining trends in thermo-mechanical properties observed from rheological and mechanics experiments \cite{knorr2012}. In particular, the mechanical properties have been investigated using atomistic models in non-equilibrium molecular dynamics simulations \cite{Sirk.2013, knorr2015}. In such studies, either the strain rate employed for uniaxial deformation or the frequency used for oscillatory deformation is far separated from the values used in experiments \cite{sirk2016bi}. Despite significant advancements in computational resources, there is a vast timescale disparity, at least six orders of magnitude, at which mechanical properties can be probed in simulations compared to experiments. Such a huge gap in timescale often lead to discrepancies in the measurement of properties. However, recent theoretical developments in non-affine lattice dynamics (NALD), an approach centered on solving the equations of motion for the non-affine displacement of a tagged particle in a disordered environment, has the potential to bridge the gap between numerical and experimental observations.

Many techniques have been proposed in the recent literature to bridge the timescale gap, incorporating accelerated atomistic dynamics to reach closer to experimentally accessible timescales. Such approaches are based on activation-relaxation technique \cite{barkema1996event}, hyperdynamics \cite{voter1997hyperdynamics}, dimer bond method \cite{henkelman1999dimer}, temperature accelerated dynamics \cite{so2000temperature}, autonomous basin climbing method \cite{kushima2009computing, zhang2021strain}, parallel replica dynamics \cite{perez2015parallel}. Also, on the experimental side, attempts have been made to realise extreme conditions to reach lower limits of numerical simulations \cite{lee2023polyurethane}. All such timescale bridging should take into account the proper modeling of bond breaking \cite{yu2023machine}, conversion of plastic work to heat, leading to temperature rise \cite{rittel1999conversion}, and frequency-dependent viscoelastic damping \cite{ranganathan2017commonalities}.

The NALD approach uses system-specific microscopic inputs, mainly the nature of the potential energy surface, to predict the frequency-dependent mechanical response of the system. At its core lies the fundamental concept of non-affine displacements. When an external deformation is applied onto a material sample, each atom tends to follow the applied strain. Then, one defines affine displacement as the atomic displacement from the original position of the atom in the undeformed sample to the position prescribed by the macroscopic strain. Since, in a disordered material, every atom is not a center of inversion symmetry, the sum of all forces acting on the atom in the affine position (due to its interactions with its neighbours, that are also moving) is not zero \cite{zaccone2023}. This is different from the situation in a perfect centrosymmetric crystal at zero temperature where the resultant of all forces in the affine position would be identically zero thanks to centrosymmetry \cite{born1996dynamical}. Hence, in a disordered material, the net force acting on the atom in the affine position has to be re-equilibrated, to maintain mechanical equilibrium, through an additional displacement, on top of the affine displacement. It is precisely this additional displacement which constitutes the \emph{non-affine} displacement. By accounting for the fact that the material has to do internal work to relax the forces via the non-affine displacements, it is possible to mathematically demonstrate that the non-affine displacements always contribute a negative (softening) term to the shear modulus \cite{zaccone2011approximate}, and the low-frequency shear modulus is thus given as $G=G_A - G_{NA}$, where $G_A$ is the affine or Born-Huang contribution, while $G_{NA}$ is the non-affine contribution.

Since the non-affine displacements can be expanded in the Hilbert space of the normal modes of the solid, the Hessian matrix or dynamical matrix plays a key role.
In particular, the diagonalization of the dynamical matrix constructed for a system configuration provides a vibrational density of states (vDOS) and eigenmodes (correlated with an affine force field), which are crucial for understanding and computing the non-affine dynamics. Such an approach, at various levels, has been successfully validated against numerical simulation data in coarse-grained models of polymer glass, in amorphous silica and thermoset polymers, in metallic glasses and in non-centrosymmetric crystals such as $\alpha$-quartz \cite{theodorou1986atomistic, lemaitre2006sum, zaccone2013disorder, milkus2016local, damart2017theory, cui2017atomic, milkus2017atomic, ness2017nonmonotonic, palyulin2018parameter, kriuchevskyi2020scaling, kriuchevskyi2022predicting,Cui_quartz}. We have already demonstrated that the NALD technique may become even more powerful if used together with the kernel polynomial method (KPM), which makes it possible to study the mechanical response of systems with a number of particles on the order of $10^5$ \cite{kriuchevskyi2020scaling}. The initial theoretical framework developed for non-affine dynamics was applied to an athermal scenario where the system always resides close to the minimum of the potential energy surface, maintaining the condition of mechanical equilibrium for each atom. Hence, the behavior of the system is fully described by standard normal modes. The framework was extended to finite temperature systems \cite{palyulin2018parameter}, taking into account the instantaneous normal modes (INMs) \cite{stratt1995instantaneous, keyes1994unstable,Douglas2019}. The INMs represent negative eigenvalues of the Hessian matrix, resulting in imaginary frequencies physically associated with the relaxation of internal tensions or internal stress \cite{palyulin2018parameter,Douglas2019}.
In such cases, finite temperature configurations from independent trajectories are used instead of energy-minimized athermal configurations.

In this work, we study the viscoelastic response of a fully cross-linked epoxy system using NALD theory at different temperatures below the glass transition temperature of the system. We use epoxy configurations which have been independently cooled from a high temperature rubber state to the target temperature, to obtain the vDOS and affine force field correlators which leads to the calculation of storage and loss modulus as a function of frequency. In the high-frequency regime, the calculations show an excellent match with the non-equilibrium molecular dynamics (NEMD) simulation for oscillatory shear. NALD calculations for storage modulus show a plateau below a certain frequency ($\approx 10^{11}$ Hz), the plateau value is found to be very close to the experimental data. This value is very sensitive to the low-frequency contributions from vDOS and affine force field correlator. Furthermore, we also show the temperature dependence of low-frequency storage modulus. This follows a similar trend as the experimental data. 

\section{Theory, Simulation, and Experiment}

\begin{figure*}[ht!]
    \includegraphics[width=15cm]{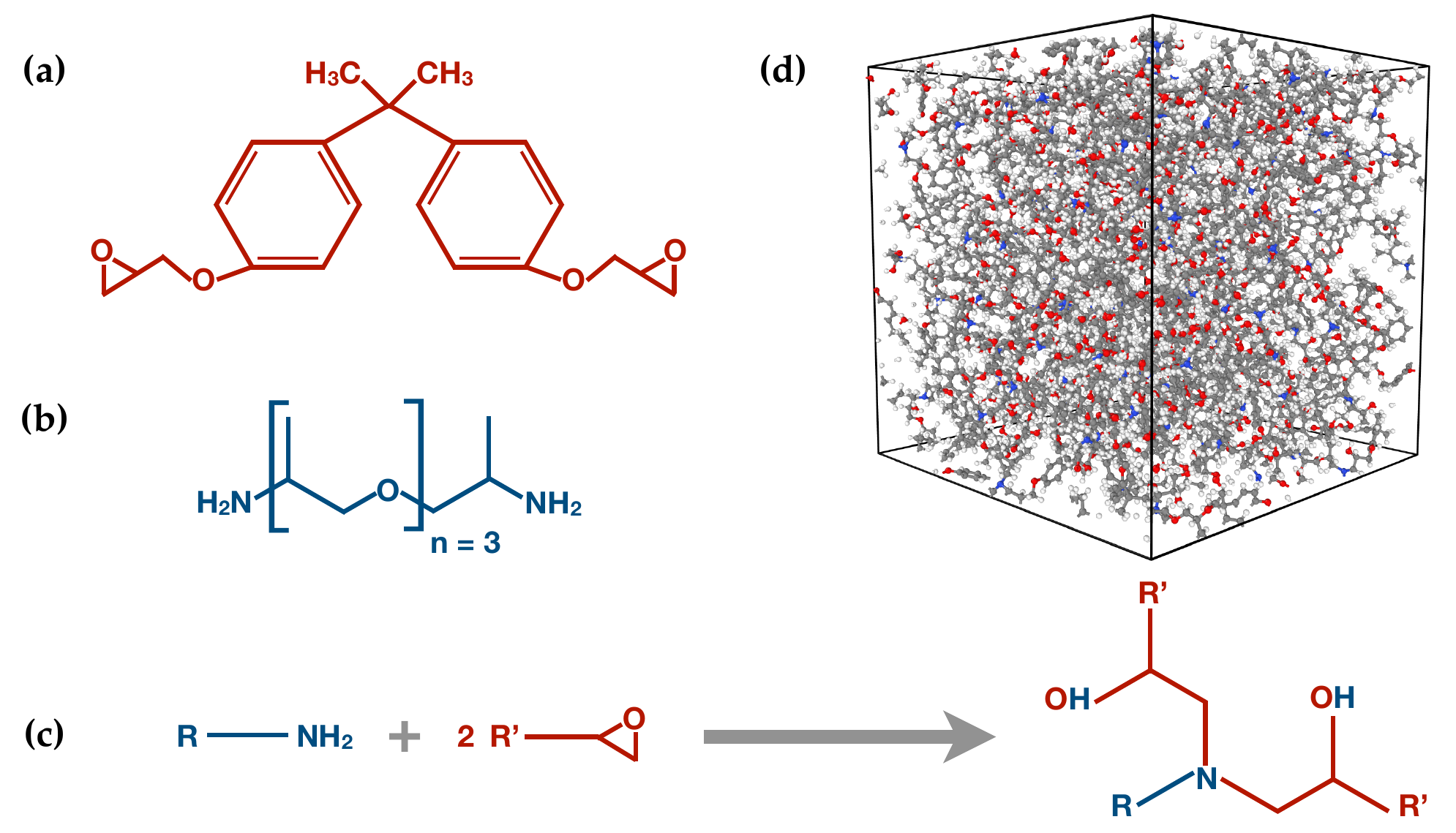}
    \caption{(a) Structure of Diglycidyl ether of bisphenol A (DGEBA). (b) Structure of poly(oxypropylene) diamine. (c) Schematic of the cross-linking reaction. (d) Snapshot of the cross-linked epoxy system.}
    \label{fig1}
\end{figure*}

\subsection{Summary of non-affine lattice dynamics}

The elastic properties of idealized centrosymmetric crystals are simplified due to the lattice inversion symmetry. In such crystals, under deformation, the displacements of atoms are affine and the local mechanical equilibrium is always satisfied \cite{born1996dynamical}. But in contrast, such a symmetry is absent in amorphous materials and the structure is largely disordered. So, in the deformed state, there is non-zero force on atoms which leads to additional (non-affine) displacements on top of the affine displacements. In this process, the stress is relaxed and the elastic free energy of the system also decreases due to the internal work. In the generalized Langevin equation framework \cite{zwanzig2001nonequilibrium}, the equation of motion of a tagged particle $i$ of mass $m$ in a deformed system can be written similar to a damped harmonic oscillator equation \cite{lemaitre2006sum, cui2017atomic}:

\begin{equation}
    m \ddot{{\bf r}}_i + \nu \dot{{\bf r}}_i + \sum_{j} {\bf H}_{ij}{\bf r}_j = {\bf \Xi}_{i}\gamma,
    \label{eom}
\end{equation}
where the effects of temperature are encoded in the INMs present in the Hessian matrix ${\bf H}_{ij}$. This equation can also be derived as a Generalized Langevin Equation (GLE) from a Caldeira-Leggett particle-bath Hamiltonian \cite{zaccone2023}, and one then notes that the stochastic force amounts to zero given its zero-average property, over a sufficient number of oscillation cycles \cite{damart2017theory}.

The second term on the left-hand side in the above equation describes a viscous damping with the dissipation coefficient $\nu$, third term is a harmonic restoring force (incorporated via the Hessian matrix ${\bf H}$) acting due to the displacement of neighboring particles and the term on right hand side is the force responsible for non-affine motion under the shear-strain $\gamma$. For a three dimensional system of $N$ particles, ${\bf H}$ is a $3N \times 3N$ matrix defined via the second derivative of the potential energy. The solution of equation (\ref{eom}), obtained by Fourier transform and normal mode decomposition, describes the deformation frequency $(\Omega)$ dependent viscoelastic modulus:

\begin{equation}
    G^*(\Omega) = G_A - \frac{1}{V} \sum_{k} \frac{\Gamma(\omega_k)}{m(\omega_k^2 -\Omega^2) + \iota \Omega \nu}.
    \label{modulus_eq}
\end{equation}

Here, $G_A$ is the contribution to the modulus due to the affine part of the motion of atoms \cite{born1996dynamical}, $V$ is the volume of the system, $\omega_k$ are the normal modes of the system and $\Gamma(\omega_k) = \tilde{{\bf \Xi}}^2_k$ (known as affine force field correlator) is the square of the projection of the force field ${\bf \Xi}$ on the eigenvector corresponding to $\omega_k$. This formalism has been generalised for multi-mass system where mass reduced affine force correlator $\tilde{\Gamma}$ is calculated \cite{kriuchevskyi2020scaling}. Hence, the ingredient to obtain the complex viscoelastic shear modulus in equation (\ref{modulus_eq}) is to obtain the Hessian matrix for the instantaneous configuration of the system and diagonalize it to get the normal mode frequencies $\omega_k$ and eigenvectors, then calculate $\tilde{\Gamma}_k$ using these vectors and force field ${\bf \Xi}$. The summation should be performed over the positive part of the real and imaginary axis. Equation (\ref{modulus_eq}) can be transformed to a continuous form using the normalized density of states $D(\omega)$

\begin{equation}
    G^*(\Omega) = G_A - \frac{3N}{V} \int \frac{D(\omega)\Gamma(\omega)}{m(\omega^2-\Omega^2) + \iota \Omega \nu}~d\omega.
    \label{modulus_eq_int}
\end{equation}

It should be noted that, although equation (\ref{modulus_eq}) and (\ref{modulus_eq_int}) are equivalent, it is useful to implement the discretized formula, as $D(\omega)$ can exhibit significant finite-size effects in small systems due to binning in $\omega$. In this work, the presence of long-range Coulomb interactions makes the matrix very dense, with each matrix requiring a few gigabytes of memory. The diagonalization of the matrix is performed using the LAPACK routines implemented in Intel MKL \cite{anderson1999lapack}. To calculate the affine modulus $G_A$, the system is affinely deformed by a very small shear-strain $\eta_{\alpha\beta} = 10^{-6}$ and the corresponding shear-stress $\sigma_{\alpha\beta}$ is measured then $G_A = \sigma_{\alpha\beta}/\eta_{\alpha\beta}$. Similarly, the force field ${\bf \Xi_{\alpha\beta}}$ responsible for non-affine motion in sheared system is calculated as the net force acting on each atom after system is affinely deformed by the same shear-strain $\eta_{\alpha\beta}$ i.e., ${\bf \Xi}_{\alpha\beta,i} = \Delta {\bf f}_i/\eta_{\alpha\beta}$, where $\Delta {\bf f}_i$ is the net force on atom $i$ with respect to the undeformed state. Mass reduced affine force correlator $\tilde{\Gamma}$ is calculated as $\tilde{\Gamma}(\omega) = (\sum_{i} {\bf \Xi}_{i} \cdot {\bf e}_{i}(\omega)/\sqrt{m_i})^2$. Here, $m_i$ is the mass of atom $i$ and ${\bf e}_{i}(\omega)$ is component of the eigenvector corresponding to eigenfrequency $\omega$ and atom $i$.

\subsection{Molecular dynamics simulations}
Our cross-linked epoxy system consists of the monomer of diglycidyl ether of Bisphenol A (DGEBA) and cross-linker polypropylene glycol (POP3). A stoichiometric reaction mixture of monomer and cross-linker molecules was equilibrated at $1$ atm and $730$ K. The monomer DGEBA (Fig.~\ref{fig1}a) and cross-linker POP3 chain ($n=3$) (Fig.~\ref{fig1}b) were cross-linked to form an epoxy-amine network as shown in Fig.~\ref{fig1}c. Cross-link bonds were assigned between epoxide and amine groups using a Monte Carlo simulated annealing approach \cite{khare1993generation, lin2009molecular}, which we have previously used to build various epoxy networks. In short, a stoichiometric number of cross-link bonds was assigned to a mixture of DGEBA-D230 molecules, such that the total length of the newly created bonds was minimized. The new bonds were thermally relaxed to their equilibrium length, excess hydrogen atoms were removed, and force field parameters were updated. A constraint on the connectivity prevented first-order loops, i.e., two bonds between a cross-linker and a single linear segment were not permitted. More details on constructing similar epoxy thermoset networks are available elsewhere \cite{Sirk.2013, sirk2016bi}. Five independent replica structures with different network connectivity were created.

Following previous studies of the same model, we apply the general AMBER force field (GAFF) \cite{cornell1995second, case2006amber}. Electrostatic interactions are computed with the particle-particle particle-mesh (PPPM) technique \cite{hockney2021computer} with a real-space cutoff at $9$ \AA. Lennard-Jones (LJ) interactions are cutoff at $9$ \AA~with tail corrections to energy and pressure measurements.\cite{sun1998compass} Molecular dynamics were carried out in the constant NPT ensemble with the LAMMPS package \cite{Plimpton.1995} (i.e., conditions at the constant number of particles, hydrostatic pressure, and temperature), where the temperature and pressure were controlled using a Nosé-Hoover thermostat and barostat \cite{shinoda2004} with temperature and pressure damping parameters of $0.1$ ps and $1$ ps, respectively, and a fixed isotropic pressure of $1$ atm. The equations of motion are integrated via the velocity-Verlet algorithm using a time-step of $1$ fs. 

Each cross-linked replica has $10074$ atoms. To obtain low temperature samples, these replicas are fully equilibrated under NPT condition at $T = 795$ K and then independently cooled step-wise, until we reach low temperature glassy states. In each step temperature is instantaneously reduced by $15$ K and followed by a NPT simulation for $2$ ns. Such a cooling protocol allows the generation of stable and relaxed states which we use for our analysis via NEMD and NALD. The resulting densities compare well with experimental values for neat epoxy matrices of $1.186$ g/cm$^3$ \cite{mcgrath2008}.

The complex modulus $G^*$ was found from non-equilibrium molecular dynamics simulations under oscillatory strain. A small amplitude oscillatory simple shear $\gamma(t)$ was applied to the epoxy configurations for which NALD analysis has been carried out, to study the response $\sigma(t)$. Here, $\gamma(t) = \gamma_0~\sin(\Omega t)$  and $\sigma(t) = \sigma_0~\sin(\Omega t+\delta)$; $\Omega$ is the frequency of deformation and $\delta$ is the phase difference between strain and stress. Using this information, the storage and loss components of the complex modulus $G*$ can be calculated as $G^{\prime} = (\sigma_0/\gamma_0) \cos\delta$ and $G^{\prime\prime} = (\sigma_0/\gamma_0) \sin\delta$. We fix $\gamma_0 = 0.01 $ and keep the thermostat active during the deformation. At least the first $50$ cycles of shear-deformation are ignored before starting the sampling of the data for the estimation of $G^{\prime}$ and $G^{\prime\prime}$.

In this work we have utilized the openly available experimental data provided by Sandia National Laboratories; the experiments were performed with a very similar epoxy system (D230) \cite{SandiaDMA}. These data have been collected in dynamic mechanical analysis (DMA) experiments, measured using ARES rheometer in torsion rectangular geometry.

\section{Results}
\subsection{Normal mode analysis}

\begin{figure}[htb!]
    \includegraphics[width=8.0cm]{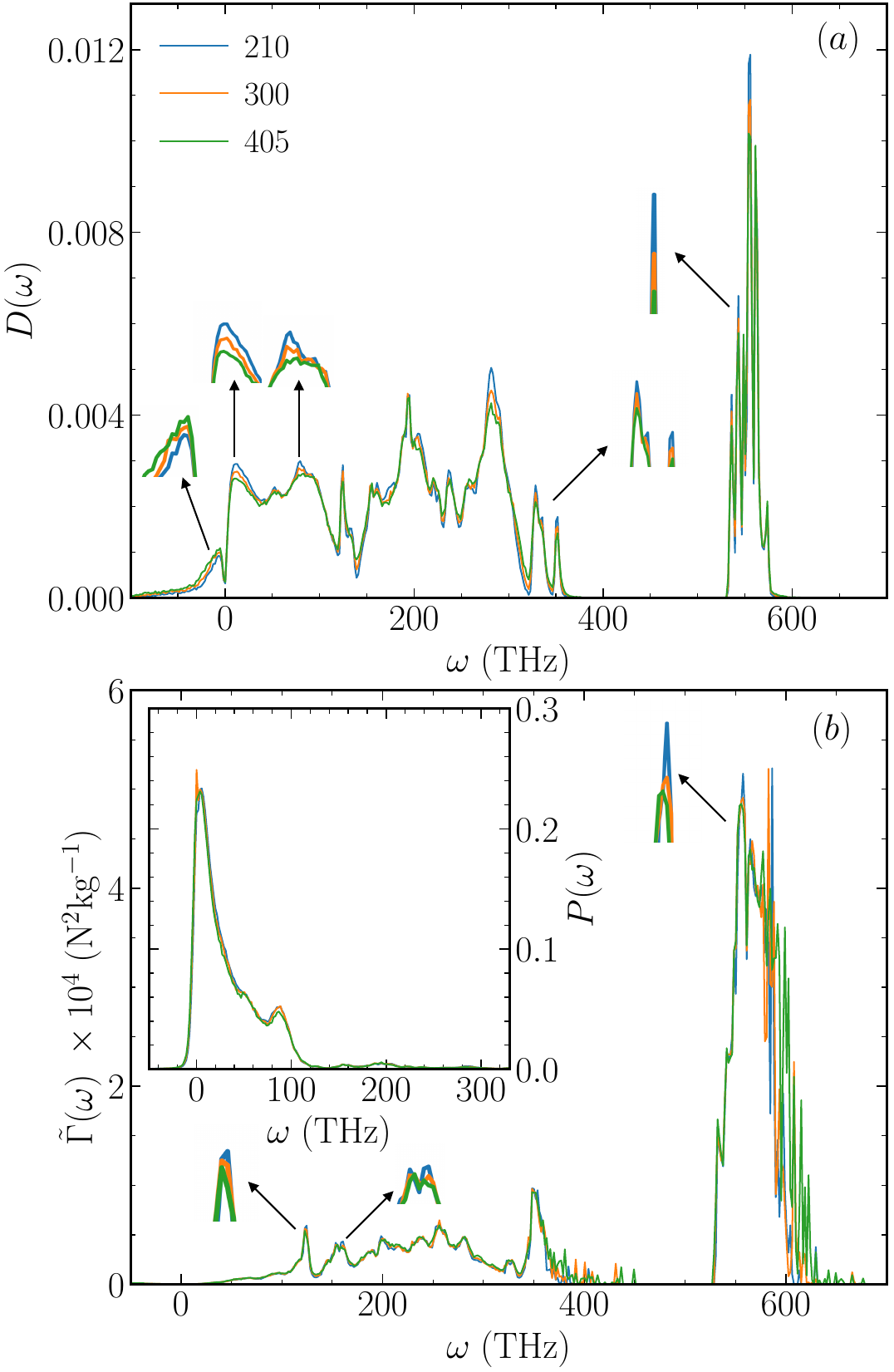}
    \caption{(a) Vibrational density of states of the system for three different temperatures $210, 300$ and $405$ K. (b) Mass rescaled affine force correlator and participation ratio (inset) as a function of frequency at different temperatures. In both the plots, some peaks are magnified for clarity.}
    \label{fig:vDOS}
\end{figure}

We compute the Hessian $H_{ij}$ using the method of finite-difference for the instantaneous configurations at different temperatures, without energy minimization. We get a significant number of negative eigenvalues which are linked with imaginary modes, indicating that the system often explores saddles in the energy landscape\cite{keyes1994unstable}. With the increase in temperature, there is a further increase in such unstable modes. The vDOS including all the modes, stable and unstable, also known as instantaneous normal modes (INMs), is shown in Fig.~\ref{fig:vDOS}(a). Following previous conventions, we show the imaginary modes on the negative side of the frequency $\omega$. We observe several sharp peaks near $550~{\rm THz}$, mainly due to hydrogen atom vibrations, a characteristic feature of organic compounds. In general, the energy of a particular bonding interaction corresponds to a feature in the high frequency range in the vDOS. There are many small and intermediate peaks below a frequency of $360~{\rm THz}$, such peaks are the molecule-specific features spanning several atoms or more. For the imaginary modes, shown on the negative side, a small peak is visible. With the change in temperature, some characteristics near peaks show a systematic change, although the change is minor. 

We have also calculated the participation ratio $P(\omega) = 1/[N \sum_i ({\bf e}_i(\omega) \cdot {\bf e}_i(\omega) )^2]$ where ${\bf e}_i(\omega)$ is the $i$th atom's component of the normalized eigenvector corresponding to mode $\omega$. This is a measure of the number of atoms participating in a single mode, where a high value of participation ratio signifies delocalization. We show the behavior of the participation ratio for our system in the inset of Fig.~\ref{fig:vDOS}(b). Participation of modes around $\omega = 0$, including some imaginary modes, is maximum and then it decreases and becomes very small near $125~{\rm THz}$. There is a small signal again near $200~{\rm THz}$ but elsewhere there is small participation. Even the modes near $550~{\rm THz}$ where vDOS shows maximum peak due to hydrogen bonding, the participation is not significant. Also, a large portion of imaginary modes, except a few unstable modes near $\omega=0$, is localized.  The participation ratio is not very sensitive across changes in temperature for the range $210-405$ K.

In Fig.~\ref{fig:vDOS}(b), we show the affine force field correlator $\tilde{\Gamma}(\omega)$ as a function of normal mode frequency $\omega$ for the three temperatures $T = 210, 300, 405$ K. Note that, since our organic system has atoms of several masses, we calculate the mass reduced correlator \cite{kriuchevskyi2020scaling}. The behavior of $\tilde{\Gamma}(\omega)$ is very similar to vDOS. There are significant peaks near $550~{\rm THz}$ as well as smaller peaks below $400~{\rm THz}$. But there are very weak but important signals for very small real and imaginary frequencies. Actually, the expression for modulus in equation (\ref{modulus_eq}) has $\omega^2$ term in the denominator, this takes over $\tilde{\Gamma}$ in the large frequency regime. But at small $\omega$, the behavior of $\tilde{\Gamma}$ dominates and significantly affects the estimates of modulus.

\subsection{Viscoelastic response}

\begin{figure}[tb!]
    \centering
    \includegraphics[width=8.0cm]{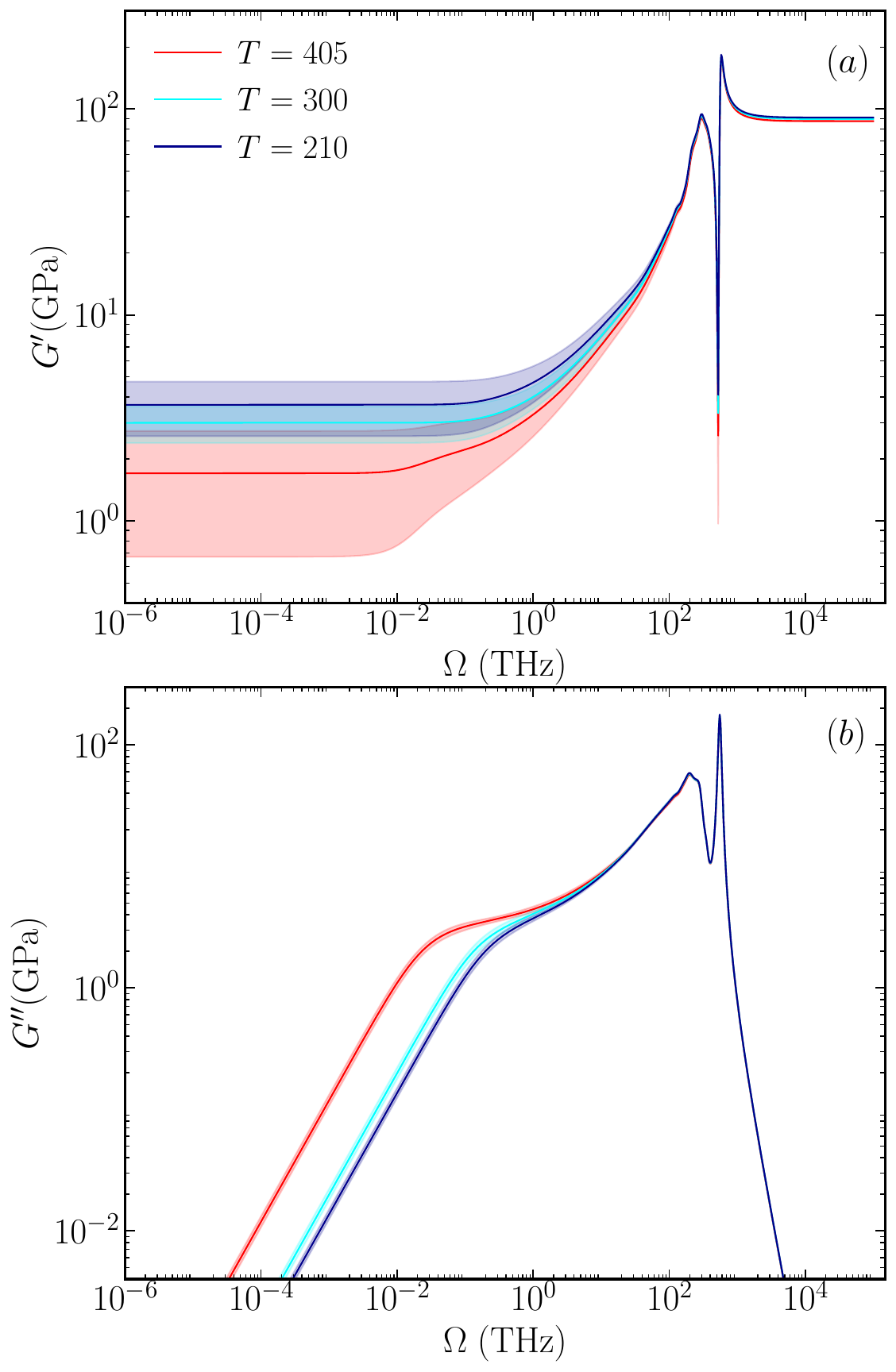}
    \caption{(a) Real part $(G^{\prime})$ and (b) imaginary part $(G^{\prime\prime})$ of the complex modulus as a function of deformation frequency $\Omega$ at different temperatures (marked). The shaded regions in both the plots indicate the error bar, calculated by the standard deviation of the measurements.}
    \label{fig:modulus_NALD}
\end{figure}

Using the framework of NALD, discussed in the previous section, we calculate the viscoelastic modulus as a function of shear deformation frequency $\Omega$ at different temperatures below the glass transition. In particular, we use equation (\ref{modulus_eq}) (but modified for multimass systems \cite{kriuchevskyi2020scaling}) to calculate the real $(G^{\prime})$ and imaginary parts $(G^{\prime\prime})$ of the complex modulus $G^* = G^{\prime} + i G^{\prime\prime}$. As also pointed out earlier, we perform the summation over the positive branch of the real and imaginary axis, taking into account the unstable modes to incorporate the thermal effects. Since, our system size is very small, so a fraction of small frequency modes are not physical to the system, because the system cannot support the propagation of the modes which correspond below the speed of sound in the medium. The smallest frequency $\omega_{\rm min}$ supported by the system can be estimated using $\omega_{\rm min} = 2\pi c_s/L$, where $c_s = \sqrt{(B+(3/4)G)/\rho}$ is the speed of sound, $L$ is the smallest dimension of the system, $\rho$ is mass density, $G$ is the magnitude of viscoelastic shear modulus and $B$ is bulk modulus of the system. Here we use the experimental value of $G$ in the estimation of cutoff frequency, with $B = (2/3)G(1+\nu)/(1-2\nu)$ and Poisson ratio $\nu = 0.35$ \cite{Sirk.2013}. Hence, all the modes smaller than the magnitude of $\omega_{\rm min}$ are ignored while evaluating the summation for viscoelastic modulus using NALD.

Fig.\ref{fig:modulus_NALD} shows $G^{\prime}$ and $G^{\prime\prime}$ (also known as storage and loss modulus respectively) as a function of external frequency $\Omega$. We observe, there are two peaks in both plots, the first one is near $200~{\rm THz}$ and the second one, which is also the highest peak, is close to $550~{\rm THz}$, corresponding to the resonating frequency range of hydrogen bonds. Between the peaks, there is a sharper decrease in $G^{\prime}$ and a minor decrease in $G^{\prime\prime}$. The value in the high $\Omega$ regime after the peaks is roughly constant for $G^{\prime}$ and close to $G_A$, the affine part of the modulus; while for $G^{\prime\prime}$ it decreases substantially. For decreasing $\Omega$ below the peaks, $G^{\prime}$ decreases slowly and reaches a temperature-dependent plateau, and $G^{\prime\prime}$ decays rapidly after the slow decay. With the change in temperature, the low $\Omega$ behavior of moduli shows sensitivity, especially the plateau value of $G^{\prime} (\Omega \to 0) = G^{\prime}_p$ increases as per the expectation, reflecting the increase in rigidity; but $G^{\prime\prime}$ shows an opposite trend that indicates the increase in dissipation.

\begin{figure}[tb!]
    \centering
    \includegraphics[width=9.0cm]{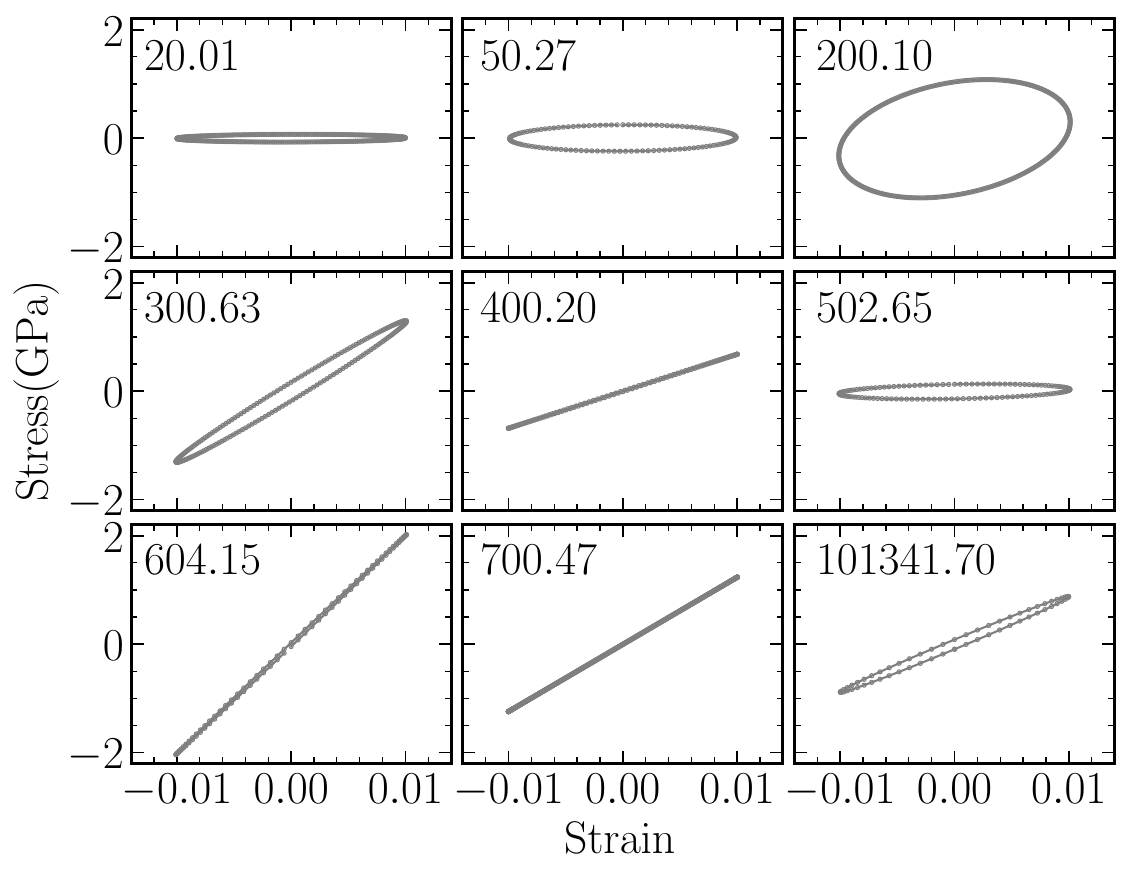}
    \includegraphics[width=9.0cm]{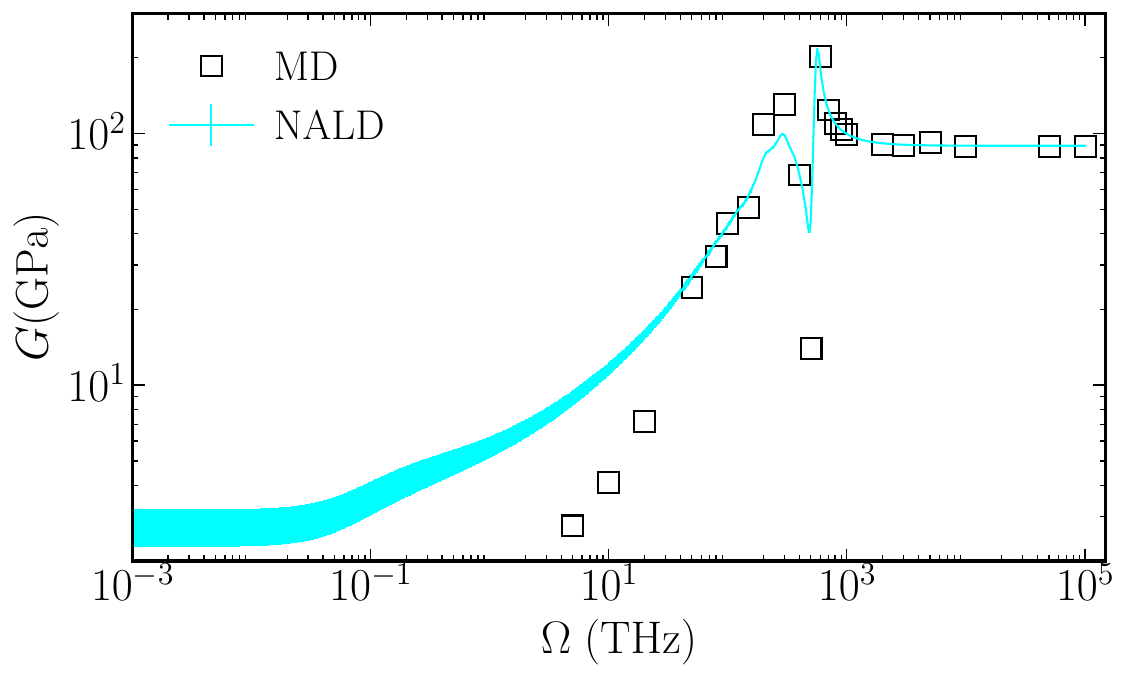}
    \caption{(top) Stress-strain relationship in the non-equilibrium molecular dynamics simulation for oscillatory shear at different frequencies $\Omega$ marked in each subplot. (bottom) Calculation of modulus $(G)$ as a function of deformation frequency $\Omega$ using MD simulation and NALD theory at temperature $T = 300$ K. The shaded region shows the error bar, calculated by the standard deviation of the measurements.}
    \label{fig:stress_strain}
\end{figure}

\section{Validation against MD data at high frequency}
Now we validate the NALD estimates of modulus against the non-equilibrium molecular dynamics simulations for oscillatory shear (mechanical spectroscopy). Following the method discussed in previous section, we shear the $xy$-plane of the simulation box with amplitude $\gamma_0 = 0.01$ and frequency $\Omega$ while measuring the stress response $\sigma_{xy}$ as a function of time at temperature $T = 300$ K. The stress-strain relationship in steady state for different $\Omega$ is shown in the top panel of Fig.~\ref{fig:stress_strain}. The relationship is very diverse across different $\Omega$. The magnitude of complex modulus $G = |G^*|$ obtained using NALD (solid line) and MD simulations (open rectangular symbols) is shown in the bottom panel of Fig.~\ref{fig:stress_strain}. We adjust the value of the NALD friction parameter $\nu$ to match the peaks near the resonating frequencies with MD values, resulting in a $\nu$ fixed as $5.6 \times 10^{13}~{\rm kg~s^{-1}}$. Overall, there is a nice match between NALD and MD estimates of modulus, except for a few data points near $10~{\rm THz}$. At lower external frequencies, possibly due to the small size and to the fact that MD becomes less reliable as the external frequency decreases, the stress of the system does not respond well \cite{rottler2003shear,sirk2016bi}. 


\begin{figure}[tb!]
    \centering
    \includegraphics[width=8.0cm]{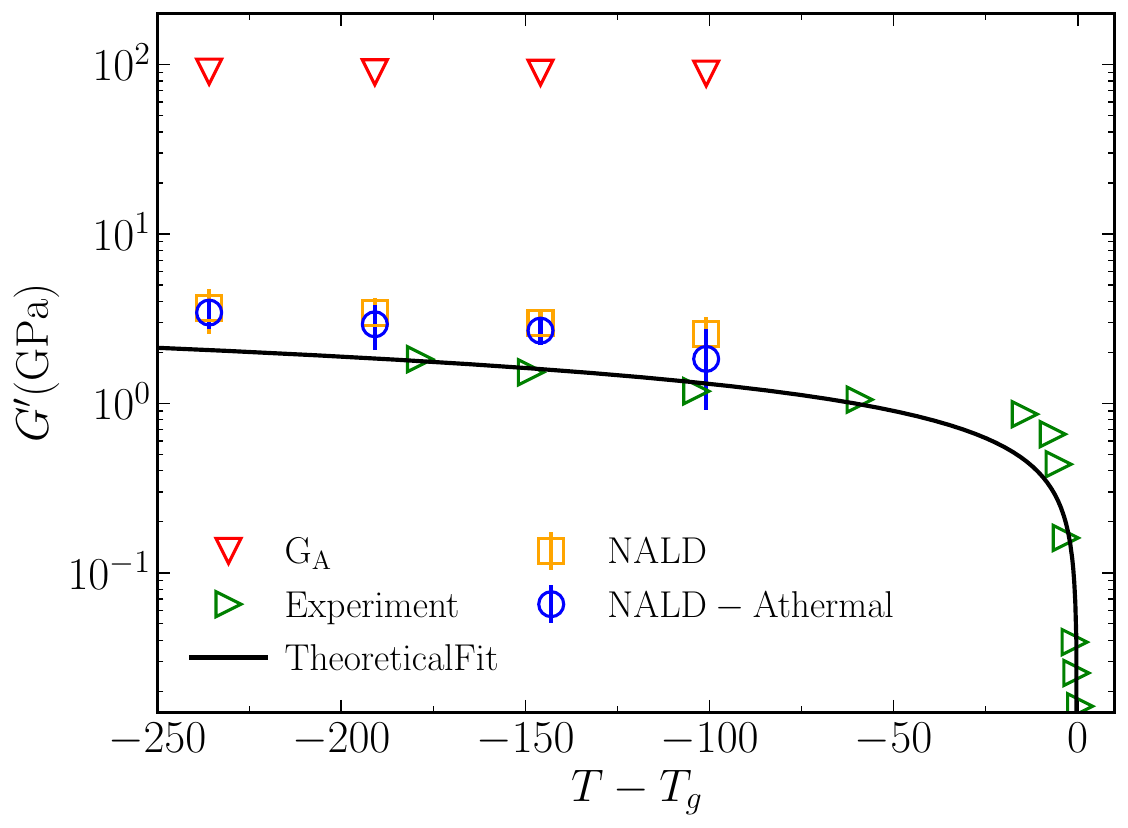}
    \caption{The plateau modulus extracted from the small frequency regime of $(G^{\prime})$ estimated by NALD in Fig.~\ref{fig:modulus_NALD}(a) (open squares) and the storage modulus estimated in DMA experiments (open right triangles) on a similar system is plotted as a function of temperature. For the athermal configurations obtained via the energy minimization of the parent configurations at high temperatures, the NALD estimate of plateau modulus is shown with open circles. Also, the temperature dependence of affine modulus $G_A$ (open down triangles) is plotted. The experimental data is fitted by the theoretical prediction (solid black line) to extract the experimental glass transition which is close to $T_g^E \approx 358.5$ K. The horizontal temperature axis has been shifted with respect to the experimental glass transition $T_g^E$ for the experimental data and with respect to the ideal glass transition $T_g^I$ of the present model simulation for the other data (see text for further details). The error bars shown with thermal and athermal measurements of NALD are calculated by the standard deviation of the measurements.}
    \label{fig:G_prime}
\end{figure}


\section{Comparison with experimental data}
We also validate the NALD estimates of plateau modulus $G^{\prime}_p$ at different temperatures against experimental data collected by dynamic mechanical analysis (DMA) of similar cross-linked epoxy networks \cite{SandiaDMA}. We show this comparison in Fig.~\ref{fig:G_prime}. Similar to experimental data, our theoretical estimates also show a decreasing behavior with the increase in temperature. There could be various reasons for the small mismatch between the theoretical and experimental data. The sample size used for the theoretical estimate is very small and the very high cooling rate used in the preparation of these samples in simulation compared to experiments would definitely lead to observational differences in properties, as can be seen in the glass transition temperature $T_g^I$ (close to $445.9$ K from Ref. \cite{Sirk.2013} for similar model system) estimated in simulation data is very different from experiments. That is why, in Fig.~\ref{fig:G_prime}, to emphasize the fact that such a difference in glass transition temperature due to the difference in thermal history can lead to a significant divergence in measurements of properties, we plot the horizontal temperature axis shifted by $T_g^I$. The sharp drop observed in modulus near experimental glass transition temperature $T_g^E$ (close to $358.5$ K, estimated below) has not been possible to capture in the NALD analysis, due to a similar reasoning. 

We can observe that the non-affine process during the deformation contributes negatively to the modulus, leading to the softening of the material which results in the overall decrease of the modulus. For further insights, we have analyzed the athermal configurations which are the energy-minimized zero-temperature configurations of their respective parent configuration at high temperature. We use the conjugate gradient technique at constant volume to minimize the energy, so that the system reaches a local minimum. As the system is dominated by the harmonic vibrations in an athermal state, the vDOS has only real modes. The NALD calculation of plateau modulus for such states with different parent temperatures is shown in Fig.~\ref{fig:G_prime}. We observe the measurements are very close to their thermal counterparts, especially at small temperatures. This indicates that the energy minimization does not alter the system configuration very much, possibly due to the constraints of the allowable configurations in the highly cross-linked network.

To the experimental data we also fit a theoretical closed-form expression for $G'(T)$ derived in \cite{zaccone2013disorder}:
\begin{equation}
    G^{\prime} = \frac{2}{5\pi}\Big(\frac{\kappa}{R_0} \phi_c e^{\alpha_T(T_g-T)} \sqrt{\phi_c[e^{\alpha_T(T_g-T)} -1]} - \frac{k_B T}{R_0^3}e^{-\alpha_TT}\Big).
    \label{G_prime_fit}
\end{equation}

Here, $\kappa = 50$ N/m is the spring constant corresponding to the C-C bond, $R_0 = 0.3$ nm is the average size of the monomer, $\alpha_T = 0.9\times 10^{-4}$ K$^{-1}$ is the thermal expansion coefficient, $\phi_c = 0.65$ is the critical packing fraction in 3D and $k_B = 1.38\times 10^{-23}$ JK$^{-1} $ is the Boltzmann constant. We consider $T_g$ as a fit parameter, resulting in $T_g^E = 358.5$ K. We have plotted the form in equation (\ref{G_prime_fit}) in Fig.~\ref{fig:G_prime}.

\begin{figure}[tb!]
    \centering
    \includegraphics[width=8.0cm]{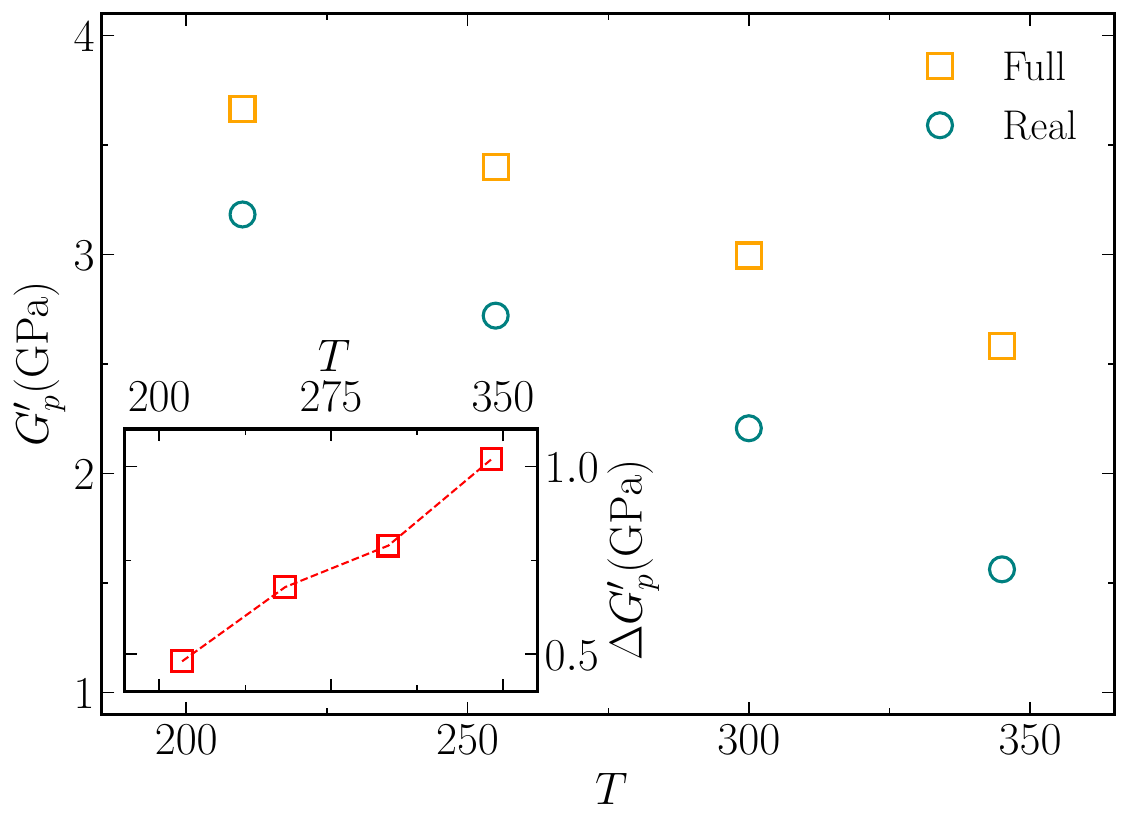}
    \caption{Temperature dependence of the NALD estimate of plateau modulus $G^{\prime}_p$ calculated by considering (open squares) and ignoring (open circles) the contributions due to the imaginary modes. The difference between the two measurements as a function of temperature is shown in the inset.}
    \label{fig:diff_plt_mod}
\end{figure}

\section{The role of instantaneous normal modes}
Finally, to understand the role of imaginary modes in deciding the plateau modulus, we have calculated the plateau modulus by considering and neglecting the contribution due to the imaginary modes. In particular, while performing the summation to calculate the modulus, first we consider the full range of $\omega$ over the real and imaginary axis (this is what has been done throughout this work) to calculate $G^{\prime}_{\rm F}$ and for the other case, we only sum over the real axis of $\omega$ to get $G^{\prime}_{\rm R}$. From these calculations, we estimate the corresponding plateau moduli  $G^{\prime}_{p \rm F}$, $G^{\prime}_{p \rm R}$ and their difference $\Delta G^{\prime}_p = G^{\prime}_{p \rm F} - G^{\prime}_{p \rm R}$ as a function of temperature, which has been shown in Fig.~\ref{fig:diff_plt_mod}. The main panel of the plot shows, $G^{\prime}_{p \rm F}$ and $G^{\prime}_{p \rm R}$ decrease with the temperature rise and at any given temperature $G^{\prime}_{p \rm F}$ is bigger than $G^{\prime}_{p \rm R}$. This means there is a positive contribution due to imaginary modes which leads to an increment in the value of modulus. The difference $\Delta G^{\prime}_p$ plotted in the inset clearly highlights that the contribution of imaginary modes increases with the increase in temperature. This is expected because, with the temperature rise, INMs have an increasing number of imaginary modes. 

Physically, this fact is due to thermal motions that instantaneously push atoms away from the harmonic-like minima towards farther separation distances in the interatomic interaction potential. These farther distances tantamount to bond tensions, or, in other words, the bonds (including van der Waals and electrostatic-type interactions) are stretched. In turn, bond-stretching is responsible for a reduction of the softening non-affine contribution, and hence leads to a larger rigidity, as was heuristically proposed by S. Alexander \cite{Alexander} for soft polymer networks, and later mathematically proved in Ref. \cite{Cui2019}.


\section{Conclusions and outlook}
We have investigated the long-standing problem of timescale bridging in the context of cross-linked epoxy polymer. Our theoretical approach using non-affine lattice dynamic (NALD) successfully bridges (a time scale gap of at least six orders of magnitude) the experimental observations of mechanics in the small frequency (large timescale) range with the simulation measurements in the high frequency (small timescale) regime. The cross-linked epoxy network prepared in molecular dynamics simulations by cooling the high-temperature rubbery state to a glassy state was analyzed for the vibrational density of states, participation ratio, and mechanical response via oscillatory shear. Thermal effects on mechanics were captured, evidenced by the measurement of the temperature dependence of plateau modulus. It was demonstrated that the non-affine softening reduces the affine modulus of the system by almost two orders of magnitude. Furthermore, we analyzed the role of imaginary modes (instantaneous normal modes) in determining the plateau modulus. A large portion of the shortcomings like slight mismatch between NALD estimate and experimental data, absence of sharp drop in modulus near $T_g^E$, etc, could, perhaps, be managed if the analysis in the present study is going to be extended to a bigger system with $\sim 10^5$ particles. The technical challenge lies in the dense Hessian matrix of such a system with the presence of long-range interactions, which poses prohibitive demands in terms of memory. Efficient management of the diagonalization of big and dense Hessian matrices will allow access to low-frequency behavior of $D(\omega)$ and $\tilde{\Gamma}(\omega)$, which are crucial for a more accurate determination of the plateau modulus at experimentally accessible frequencies and higher temperatures.

\section*{Conflicts of interest}
There are no conflicts to declare.

\section*{Acknowledgements}
A.Z. and V.V. gratefully acknowledge funding from the European Union through Horizon Europe ERC Grant number: 101043968 ``Multimech''. A.Z. gratefully acknowledges the Nieders{\"a}chsische Akademie der Wissenschaften zu G{\"o}ttingen in the frame of the Gauss Professorship program and funding from US Army Research Office through contract nr. W911NF-22-2-0256.

\bibliography{epoxy} 

\end{document}